\documentclass[aps,prd,twocolumn,groupedaddress,nofootinbib,showpacs]{revtex4}

\usepackage{euscript,graphicx,amssymb,subfigure}

\usepackage{amsfonts}
\usepackage{amsmath}
\usepackage{amssymb}
\usepackage{graphicx}
\usepackage{euscript}
\usepackage{color}
\usepackage{subfigure}

\newcommand{\goodgap}{\hspace{\subfigtopskip} \hspace{\subfigbottomskip}}

\begin{document}

\title{Testing the distance duality relation with present and future data}

\author{V.F. Cardone$^{1}$, S. Spiro$^{1}$, I. Hook$^{1,2}$, R. Scaramella$^{1}$}

\affiliation{
$^1$I.N.A.F.\,-\,Osservatorio Astronomico di Roma, via Frascati 33, 00040 - Monte Porzio Catone (Roma), Italy \\
$^2$Department of Astrophysics, University of Oxford, Denys Wilkinson Building, Keble Road, Oxford, United Kingdom}

\begin{abstract}

The assumptions that {\it light propagates along null geodesics of the spacetime metric} and {\it the number of photons is conserved along the light path} lead to the distance duality relation (DDR), $\eta = D_L(z) (1 + z)^{-2}/D_A(z) = 1$, with $D_L(z)$ and $D_A(z)$ the luminosity and angular diameter distances to a source at redshift $z$. In order to test the DDR, we follow the usual strategy comparing the angular diameter distances of a set of clusters, inferred from X\,-\,ray and radio data, with the luminosity distance at the same cluster redshift using the local regression technique to estimate $D_L(z)$ from Type Ia Supernovae (SNeIa) Hubble diagram. In order to both strengthen the constraints on the DDR and get rid of the systematics related to the unknown cluster geometry, we also investigate the possibility to use Baryon Acoustic Oscillations (BAO) to infer $D_A(z)$ from future BAO surveys. As a test case, we consider the proposed Euclid mission investigating the precision can be afforded on $\eta(z)$ from the expected SNeIa and BAO data. We find that the combination of BAO and the local regression coupled allows to reduce the errors on $\eta_a = d\eta/dz|_{z = 0}$ by a factor two if one $\eta_0 = \eta(z = 0) = 1$ is forced and future data are used. On the other hand, although the statistical error on $\eta_0$ is not significantly reduced, the constraints on this quantity will be nevertheless ameliorated thanks to the reduce impact of systematics.

\end{abstract}


\maketitle

\section{Introduction}

The Etherington reciprocity theorem \cite{E33} states that, if source and observer are in relative motion, solid angles subtended between the observer and the source are related by geometrical invariants where the redshift of the source as measured by the observer enters in the relation. First proven in the context of relativistic geometrical optics, it only relies on the two assumptions that light travels along null geodesics in a Riemannian spacetime and that the number of photons is conserved \cite{E71}. Altough often underrated, the Etherington reciprocity theorem actually plays a fundamental role in observational cosmology with applications ranging from gravitational lensing \cite{SEF}, to the CMBR temperature shift equation $T_e = T_0/(1 + z)$ \cite{E71} and the well known result that the surface brightness of a source does not depend on its distance to the observer. Among its different incarnation, a widely used formulation of the Etherington reciprocity theorem is represented by the so called {\it distance duality relation} (hereafter, DDR \cite{BK03}) reading\,:

\begin{equation}
\eta(z) = \frac{D_L(z) (1 + z)^{-2}}{D_A(z)} = 1
\label{eq: ddr}
\end{equation}
where $D_L(z)$ and $D_A(z)$ are the luminosity (LD) and angular diameter distance (ADD). Having been derived from the reciprocity law, the DDR holds in whatever cosmology provided the spacetime is Riemannian and there are no source of attenuation (like gray dust) or brightening (as gravitational lensing). As such, one can take it for granted, but a more interesting possibility is to test it against astronomical observations. To this end, one should be able to measure, for a given $z$, both the LD and ADD by means of a standard candle and a standard ruler, respectively. From this point of view, Type Ia Supernovae (although standardizable rather than standard candles) are the ideal tool to estimate the LD as is indeed routinely done when using their Hubble diagram to constrain cosmological parameters. On the contrary, ADDs are much more difficult to measure, but some significant steps forward have been recently based on the Sunyaev\,-\,Zel'dovich effect in galaxy clusters \cite{Betty,B06}. Unfortunately, while the method to estimate ADD from the measured temperature decrement is theoretically and observationally well understood, the impact of systematics related to the cluster geometry and the plasma physics is still quite strong leading to contrasting conclusions on the DDR validity at any redshift \cite{DDRvsObs,DDRvsCl}.

As a further issue, one has also to take care of the errors due to the mismatch between the cluster redshift and the closest SN in the companion SNeIa sample adopted. Different strategies have been implemented to avoid this problem (e.g., by rejecting the clusters for which no SN at the same $z$ is available) or reduce its impact relying on the LD value inferred from SNeIa with $|\Delta z| \le 0.005$ and $\Delta z = z_{SN} - z_{cl}$. As a possible way out of this issue, we present here a novel method relying on the {\it local regression} technique \cite{LR} to get a reliable LD estimate at exactly the same redshift as the cluster one.

An alternative standard ruler is represented by the sound horizon $r_s$, i.e. the comoving distance a sound wave could have traveled in a photon\,-\,baryon fluid by the time of decoupling. The importance of such a scale may be guessed noting that, at the time of recombination, baryons wave stop to freely propagate in the initial baryons\,-\,photons plasma thus leaving a density excess at the sound horizon scale. Should galaxy form at the centre of density perturbations, we should have observed a peak in the galaxy correlation function at this scale. Since the Fourier transform of such a peak would appear as an oscillating feature, the matter power spectrum should present oscillations at the corresponding wavenumber. Such oscillations have been indeed detected \cite{Eis05,P10} and are now referred to as {\it Baryon Acoustic Oscillations} (BAO, see \cite{BH09} for a nice review). Should one be able to measure the power spectrum as function of both the parallel and transverse wave number at different redshift $z$, BAO would allow to determine the values of $r_s H(z)$ and $D_A(z)/r_s$, where $H(z)$ is the Hubble expansion rate. Although BAO data actually determine ADDs only up to the unknown sound horizon $r_s$, it is worth noting that this latter quantity is well constrained by present day CMBR data with a precision which will likely increase as the Planck mission data \cite{Planck} will become available. Moreover, the inferred ADDs from BAO and the CMBR determination of $r_s$ will be free of the unknown systematics related to the cluster geometry and physics. We will therefore investigate here whether future BAO and SNeIa surveys can be combined together to strengthen the constrains on $\eta(z)$ and detect any DDR violation.

The plan of the paper is as follows. The local regression technique is presented in Section II and then used to infer the $\eta(z)$ values from the present day SNeIa and cluster data. Section III investigates the constraints these data put on two different parameterizations of $\eta(z)$ highlighting to what extent they depend on the cluster geometry assumptions. The use of BAO as alternative standard rulers is presented in Section IV, where we also investigate the constraints this method can impose on $\eta(z)$ relying on future SNeIa and BAO data which will be collected by the Euclid satellite. We then summarize and conclude in Section V.

\section{Avoiding redshift mismatch}

Since the DDR involves the ratio between the values of the LD scaled by $(1 + z)^{-2}$ and the ADD at the same redshift $z$, the first issue one has to tackle off is the difficulty to exactly match the measurements of these two quantities. As an example, let us consider the ADD catalog assembled by Bonamente et al. (hereafter, B06) \cite{B06} from 38 galaxy clusters spanning the redshift range $(0.149, 0.890)$. To trace the LD, we will use the most updated SNeIa sample, namely the Union2 \cite{Union2}, with 557 SNeIa over the range $(0.29, 1.40)$. Should we decide to only use the LD and ADD measurements with exactly the same $z$ in the two catalogs, we would have obtained a sample of only 13 objects with large error bars so that the results on testing the DDR would likely be quite poor.

In an attempt to strengthen the constraints, one therefore adopt an approximate matching by selecting only those clusters which have at least one SN with $|\Delta z| \le \Delta_{max}$. For $\Delta_{max} = 0.001$ ($0.005$), one finds 32 (38) objects and then estimate the LD at $z_{cl}$ from the sample of LD measurements approximately matched for each $z_{cl}$. Two strategies are possible to this end. First, one can simply take a weighted mean (with the inverse squared error as weights) or linearly interpolate the data. As we will show later, the choice of the LD estimate method and the value of $\Delta_{max}$ have a non negligible impact on the constraints on the DDR parameters. In order to reduce this bias, one should make $\Delta_{max}$ as small as possible, but this comes at the price of weakening the constraints so that finding the right compromise is an hard issue.

As a possible way out of this problem, we resort here to the {\it local regression} (LR) technique \cite{LR} to infer the distance modulus $\mu$ at the cluster redshift $z_{cl}$ from the companion SNeIa sample. The basic idea underlying LR relies on fitting simple models to localized subsets of the data to build up a function that describes the deterministic part of the variation in the data, point by point. Actually, one is not required to specify a global function of any form to fit a model to the data so that there is no ambiguity in the choice of the interpolating function. Indeed, at each point, a low degree polynomial is fit to a subset of the data containing only those points which are nearest to the one whose response is being estimated. The polynomial is fit using weighted least squares with a weight function which quickly decreases with the distance from the point where the model has to be recovered. We use the Union2 SNeIa sample as input to the local regression estimate of $\mu(z)$ following the steps schematically sketched below.

\begin{enumerate}

\item{Order the SNeIa according to increasing value of $|z_{cl} - z_i|$ and select the first $n = \alpha {\cal{N}}_{SNeIa}$ with $\alpha$ a user selected value and ${\cal{N}}_{SNeIa}$ the total number of SNeIa. \\}

\item{Define the weight function\,:

\begin{equation}
W(u) = \left \{
\begin{array}{ll}
(1 - |u|^3)^3 & |u| \le 1 \\ ~ & ~ \\ 0 & |u| \ge 1
\end{array}
\right .
\label{eq: wdef}
\end{equation}
where $u = |z_{cl} - z_i|/\Delta$ and $\Delta$ the maximum value of the $|z_{cl} - z_i|$ over the subset chosen before. \\ }

\item{Fit a first order polynomial to the data selected at step (ii) weighting each SNeIa with the corresponding value of the function $W(u)$ and take the zeroth order term as best estimate of $\mu(z)$. \\ }

\item{Estimate the error on $\mu(z)$ as the root mean square of the weighted residuals with respect to the best fit zeroth order term. \\ }

\end{enumerate}
It is worth stressing that both the choice of the weight function and the order of the fitting polynomial are somewhat arbitrary. Similarly, the value of $\alpha$ to be used must not be too small in order to make up a statistical valuable sample, but also not too large to prevent the use of a low order polynomial. In \cite{CCD09} (which we refer the reader to for any detail), an extensive set of simulations were performed to both check the reliability of the LR method and look for a possible value of $\alpha$. It was there shown that setting $\alpha = 0.025$ allows to recover the input distance modulus typically within $0.35\%$ (and with deviations never larger than $1\%$) independent on the redshift $z$ and the cosmological model adopted (at least within the large class of dark energy equation of states considered). We will therefore adopt the above procedure to estimate the distance modulus and then the LD, $D_L(z) = dex[(\mu - 25)/5]$ (with $dex(x) = 10^x$) for all the clusters in the ADD catalogs we will use later.

\section{DDR vs present day data}

Testing the validity of the DDR is the same as checking that the parameter $\eta(z)$ defined in Eq.(\ref{eq: ddr}) is strictly constant and unity at all $z$. To this end, it is convenient to phenomenologically parameterize this quantity so that deviations from the validity of the DDR can be expressed in a quantitative way. Inspired by the analogy with the dark energy equation of state, two common expressions adopted in literature read \cite{DDRvsObs, DDRvsCl}\,:

\begin{equation}
\eta(z) = \left \{
\begin{array}{l}
\eta_0 +\ \eta_a z/(1 + z) \\
~ \\
\eta_0 +\ \eta_a \ln{(1 + z)} \\
\end{array}
\right . \ ,
\label{eq: etapar}
\end{equation}
so that the DDR is never violated if $(\eta_0, \eta_a) = (1, 0)$. It is worth noting that, while the first formula predicts that $\eta(z)$ asymptotically approaches the constant value $\eta_0 + \eta_a$ at high $z$ (so that one can formally have a violation of DDR at low redshift but recover it for $z \longrightarrow \infty$ if $\eta_0 + \eta_a = 1)$, the second expression formally diverges at infinity so that it must be considered as a low $z$ approximation only. We nevertheless include it both to compare our results with previous ones and to allow for a quickly varying $\eta(z)$ (noting that, for the same $\eta_a$, the logarithmic ansatz increases faster than the first expression).

As a second remark, it is worth spending some words on the value of $\eta_0$. If one assumes that the Robertson\,-\,Walker metric holds (i.e., the universe is homogenous and isotropic on large scales), one gets $D_L(z = 0) = D_A(z = 0)$ and hence $\eta_0 = 1$ independent on whether the DDR holds or not. However, such a result breaks down if photons are absorbed or emitted along their light path or, put in other words, the effective opacity \cite{Verde} of the universe is not zero. In such a case, one can still have a homogenous and isotropic universe and nevertheless a value of $\eta_0 \ne 1$ so that we will explore both one parameter models forcing $\eta_0 = 1$ and two parameters cases constraining its value from the fit to the data.

The two expressions in Eq.(\ref{eq: etapar}) provide a purely phenomenological approach to testing the DDR. As a different method, it is also possible to assume a model for the absorption and/or production of photons due to interactions with, e.g., axion\,-\,like particles or a work out a different mechanism leading to a non vanishing and redshift dependent effective opacity (see, e.g., \cite{Verde} and refs. therein for some interesting examples). The price to pay is, however, to introduce a dependence of the fitting results on both the underlying cosmological model and the opacity production phenomenon parameters. Since the number and quality of the present day data is far from being good enough, we have here preferred to adopt a model independent approach relying on the above two phenomenological expressions.

As input dataset, we follow the common approach using the Union2 SNeIa sample as LD tracer and two different galaxy cluster samples with X\,-\,ray and SZ data to measure the ADDs. The first one is the catalog of 25 clusters assembled by De Filippis et al. (\cite{Betty}, hereafter DeF05), while the second one is made out 38 clusters and will be referred to here as the B06 \cite{B06} sample. It is worth stressing that, although the data and the method used to determine the ADD of each cluster are the same, the two samples differ for a critical assumption. Indeed, while B06 assumes a spherical geometry, DeF05 explicitly correct their estimates taking care of their constraints on the ellipsoidal cluster geometry. As amply discussed in literature \cite{DDRvsCl}, the assumption of a spherical or ellipsoidal geometry has a great impact on the ADD determination so that the estimated $\eta$ values are not consistent with each other. As a consequence, the constraints on $(\eta_0, \eta_a)$ will also depend on which sample is used and cannot be straightforwardly compared.

In order to constrain the parameters, we resort to the usual $\chi^2$ analysis, i.e., we minimize the merit function\,:

\begin{equation}
\chi^2({\bf p}) = \sum_{i = 1}^{{\cal{N}}}{\left [ \frac{\eta_{obs}(z_i) (1 + \Delta_0) - \eta_{th}(z_i, {\bf p})}{\sigma_i} \right ]^2}
\label{eq: defchi}
\end{equation}
with $\eta_{obs}(z_i)$ and $\eta_{th}(z_i, {\bf p})$ the observed and theoretically predicted $\eta(z)$ value at redshift $z_i$, $\sigma_i$ the measurement uncertainty and ${\bf p} = \eta_a$ or ${\bf p} = (\eta_0, \eta_a)$ for the one and two parameters models, respectively. Eq.(\ref{eq: defchi}) contains an additional term $(1 + \Delta_0)$ which we have introduced to take into account a systematic uncertainty on the LD as inferred from the SN distance modulus. Indeed, since the absolute SN magnitude is known up to $\pm 0.05 \ {\rm mag}$, the LD can be shifted by a factor $\Delta_0 \simeq \pm 2.3 \%$. We therefore add this as a nuisance parameter and marginalize over it with a Gaussian prior centred on $\langle \Delta_0 \rangle = 0$ and with standard deviation $\sigma_0 = 0.023$. As a far as we know, this is the first time such a term is taken into account\footnote{We thank the anonymous referee for suggesting its inclusion.}, while neglecting it can artificially reduce the uncertainties on the inferred constraints on the model parameters ${\bf p}$.

The best fit parameters will be obtained by minimizing the $\chi^2$ merit function, while the $68\%$ ($95\%$) confidence limits will be found by imposing $\Delta \chi^2 = 1.0$ $(\Delta \chi^2 = 4.0)$. To this end, we first integrate the likelihood ${\cal{L}}(\eta_0, \eta_a, \Delta_0) \propto \exp{[ -\chi^2(\eta_0, \eta_a, \Delta_0)/2 ]} \exp{-[\Delta_0^2/(2 \sigma_0^2)]}$ over all the parameters but the one of interest. We then define $\tilde{\chi}^2_i = - 2 \ln{{\cal{L}}_i}$ (with ${\cal{L}}_i$ the marginalized likelihood for the i\,-\,th parameter) and find the $68\%$ and $95\%$ CL solving the above relations for $\Delta \tilde{\chi}^2$.

\subsection{Taking care of redshift mismatch}

Before discussing the results on $(\eta_0, \eta_a)$ from fitting the above dataset, it is worth spending some time to explicitly show the impact of redshift mismatch and why we advice the reader to avoid it using the local regression technique (or a whatever reliable method to estimate the LD at the same cluster redshift).

To this aim, we build up simulated cluster and SNeIa samples as close as possible to the actual ones. First, we choose a fiducial cosmological model assuming a spatially flat universe with matter density parameter $\Omega_M = 0.27$, constant dark energy equation of state, $w = -0.95$ and Hubble constant (in units of $100 \ km/s/Mpc$) $h = 0.703$, consistent with the recent WMAP7 \cite{WMAP7} results. We then choose the B06 sample as a reference case and assign to each cluster in this sample an ADD equals to $\kappa D_A(z)$ with $D_A(z)$ the theoretical value and $\kappa$ randomly chosen between $(0.98, 1.02)$ to mimic a possible mismatch due to statistical and/or systematic errors. To each value, we then attach a measurement uncertainty in such a way that the relative error equals the one for the ADD of the cluster in the B06 sample having the same $z$. For the simulated SNeIa sample, we adopt a similar procedure the only difference being that we generate the distance modulus (rather than the LD) from a Gaussian distribution centred on the theoretical value and with variance $\sigma_\mu = (\mu_{sim}/\mu_{obs}) \sigma_{obs}$, but never smaller than $\sigma_{int} = 0.15$, this value being the intrinsic scatter of the SNeIa peak magnitude. The same scaling of the errors is then used to assign a statistical uncertainty to the simulated $\mu(z)$ for each SN in the sample.

The simulated cluster and SNeIa datasets are then used to estimate $\eta(z)$ at the cluster redshifts using two different ways to deal with the problem of redshift mismatch. First, we take as LD at each $z_{cl}$ the error weighted average of the SNeIa with $|\Delta z| \le 0.005$ thus obtaining\footnote{We discuss only the results obtained fitting the first $\eta(z)$ model in Eq.(\ref{eq: etapar}), but our conclusions on the impact of redshift mismatch are qualitatively the same for the other parametrization. Moreover, we report the values obtained by a single simulation, but we have checked that they are qualitatively the same running $\sim 100$ realizations of the LD and ADD datasets.}\,:

\begin{displaymath}
\eta_a = -0.071 \pm 0.100
\end{displaymath}
when forcing $\eta_0 = 1$, and

\begin{displaymath}
\eta_0 = 0.940 \pm 0.085 \ \ , \ \ \eta_a = 0.226 \pm 0.475
\end{displaymath}
for the two parameters case with the reported errors referring to the $68\%$ confidence range\footnote{Note that the marginalized distribution are very close to Gaussian so that the $68\%$ confidence range may be taken as a $1 \sigma$ error and $95\%$ CL obtained by doubling the $1 \sigma$ uncertainty. Hereafter, we will therefore report only this estimate of the $1 \sigma$ error.}. Such a test shows that, although the value $\eta_a = 0$ is well within the $68\%$ confidence ranges, the best fit value may be severely biased if one does not force $\eta_0 = 1$. Since it is reasonable to expect that the error bars will shrink with future data, we can argue that averaging over the SNeIa with $|\Delta z| \le 0.005$ can introduce a systematic bias which is larger than the statistical uncertainty.

\begin{table*}
\begin{center}
\begin{tabular}{ccccc}
\hline Sample & $\eta_a$ & $(\eta_0, \eta_a)_{bf}$ & $\eta_0$ & $\eta_a$ \\
\hline \hline
~  & ~  & ~  & ~  & ~ \\
B06 & $-0.331 \pm 0.129$ & $(0.899, -0.192)$ & $0.915 \pm 0.078$ & $-0.195 \pm 0.311$ \\
~  & ~  & ~  & ~  & ~ \\
DeF05 & $-0.622 \pm 0.232$ & $(0.719, 0.280)$ & $0.751 \pm 0.091$ & $0.292 \pm 0.538$ \\
~  & ~  & ~  & ~  & ~ \\
\hline
~  & ~  & ~  & ~  & ~ \\
B06 & $-0.273 \pm 0.125$ & $(0.896, 0.150)$ & $0.911 \pm 0.067$ & $-0.153 \pm 0.223$ \\
~  & ~  & ~  & ~  & ~ \\
DeF05 & $-0.530 \pm 0.217$ & $(0.727, 0.210)$ & $0.758 \pm 0.097$ & $0.220 \pm 0.419$ \\
~  & ~  & ~  & ~  & ~ \\
\hline
\end{tabular}
\caption{Constraints on DDR test quantity $\eta(z)$ after marginalizing over $\Delta_0$. Columns are as follows\,: 1. cluster sample used, 2. median and $68\%$ confidence range for $\eta_a$ forcing $\eta_0 = 1.0$, 3. best fit $(\eta_0, \eta_a)$ values for the two parameter model, 4., 5. median and $68\%$ confidence ranges for $(\eta_0, \eta_a)$. Upper (lower) half of the table refers to the first (second) ansatz in Eq.(\ref{eq: etapar}).}
\label{tab: fitpar}
\end{center}
\end{table*}

Actually, averaging is only zero order approximation so that one can suppose that a linear interpolation of the $D_L(z)$ values within this range works much better. Using this approach, we find\,:

\begin{displaymath}
\eta_a = -0.163 \pm 0.080
\end{displaymath}
for the one parameter model, and

\begin{displaymath}
\eta_0 = 0.836 \pm 0.052 \ \ , \ \ \eta_a = 0.148 \pm 0.171
\end{displaymath}
when $\eta_0$ is left free. It is evident that the bias on $\eta_a$ is still present for the two parameters model. Somewhat surprisingly, the linear interpolation method has worsened rather than ameliorated the situation. Actually, this is partly a consequence of the smaller number of clusters used which makes the fit more sensible to deviations from the DDR ansatz because of statistical fluctuations. Note that the dataset only contains now 28 clusters since, for ten of them, we have too few points (less than four objects) in the $|\Delta z| \le 0.005$ SNeIa subset to define a reliable interpolation.

Finally, let us consider the results obtained using local regression to estimate $\mu(z)$ and then $D_L(z)$ for each cluster in the simulated sample. Fitting the one parameter model, we get\,:

\begin{displaymath}
\eta_a = -0.005 \pm 0.126 \ ,
\end{displaymath}
while, when $\eta_0$ is fitted too, we find\,:

\begin{displaymath}
\eta_0 = 1.002 \pm 0.100 \ \ , \ \ \eta_a = -0.022 \pm 0.291 \ \ .
\end{displaymath}
Compared to the averaging method, we clearly see that the bias on $\eta_a$ is reduced both for one and two parameter models and, as a further positive outcome, we also get a median $\eta_0$ value quite close to the input one. We can therefore safely conclude that the local regression technique does not bias the constraints on $(\eta_0, \eta_a)$ and confidently advocate its use to test the DDR avoiding any systematic error due to the redshift mismatch problem.

\subsection{Present day constraints}

Motivated by the above discussion, we now use the local regression technique to infer the LD of the clusters in the B06 and DeF05 samples using the SNeIa Union2 sample as input. We then fit the data thus obtained with the four models introduced in Section II and summarize the results in Table I. Not surprisingly, the confidence ranges are quite large so that it is not statistically possible to definitively conclude whether the DDR holds or not at any $z$. It is worth noting that a qualitatively similar conclusion is also achieved in previous works. Indeed, the constraints in Table I are fully consistent with those in \cite{DDRvsObs,DDRvsCl}, although we remark that a straightforward comparison should be avoided given the radically different approach to the redshift mismatch problem. Moreover, we have also included the term $(1 + \Delta_0)$ in Eq.(\ref{eq: defchi}) which has the double impact of introducing a degeneracy in the parameters space and enlarging the confidence ranges.

It is worth investigating how the constraints depend on the assumed $\eta(z)$ parameterization. Comparing the constraints on $\eta_a$ for both the one and two parameters models in the upper and lower half of Table I, we see that the logarithmic ansatz may be reconciled with the data only if smaller $\eta_a$ values are used. This is an expected result considering that, for the same $\eta_0$ value (as, e.g., for the one parameter case), a smaller $\eta_a$ partially compensates for the different scalings with $z$ of the two cases considered. Although somewhat expected, this result highlights the importance of choosing a reliable parameterization for $\eta(z)$ in order to better check the DDR validity at any $z$. On the contrary, what is the functional expression for $\eta(z)$ has only a minor impact on the $\eta_0$ constraints. Indeed, for a fixed sample, the $68\%$ confidence ranges are well overlapped for the two $\eta(z)$ expressions so that one could draw conclusions on $\eta_0$ in a roughly model independent way.

Table I shows that, actually, the larger impact on the constraints is due to the sample used, that is to say on the assumptions on the cluster geometry. Indeed, both for models with $\eta_0 = 1$ or left free to fit, the B06 sample give values of $\eta_a$ closer to zero than the DeF05 one. Moreover, when $\eta_0$ is free to vary, the B06 sample recovers $\eta_0 = 1$ within $2 \sigma$, while a significantly smaller value, $\eta_0 \sim 0.76$, is obtained with the DeF05 sample leading to $\eta_0 < 1$ at more than $2.7 \sigma$.  Since the SNeIa companion sample used is the same, it is likely that the difference has to be ascribed to how the ADD has been estimated from the cluster data. In particular, since $\eta_0 < 1$ has been obtained, one should argue that the LD has been underestimated or the ADD is overestimated. Investigating in details this issue is outside our aims. We only stress that the uncertainty on the cluster geometry is likely to not be reduced with improved observations being related to projection effects. As a consequence, this source of systematic error is hard to be fully taken under control also with future data.

\begin{figure*}
\centering
\subfigure{\includegraphics[width=5cm]{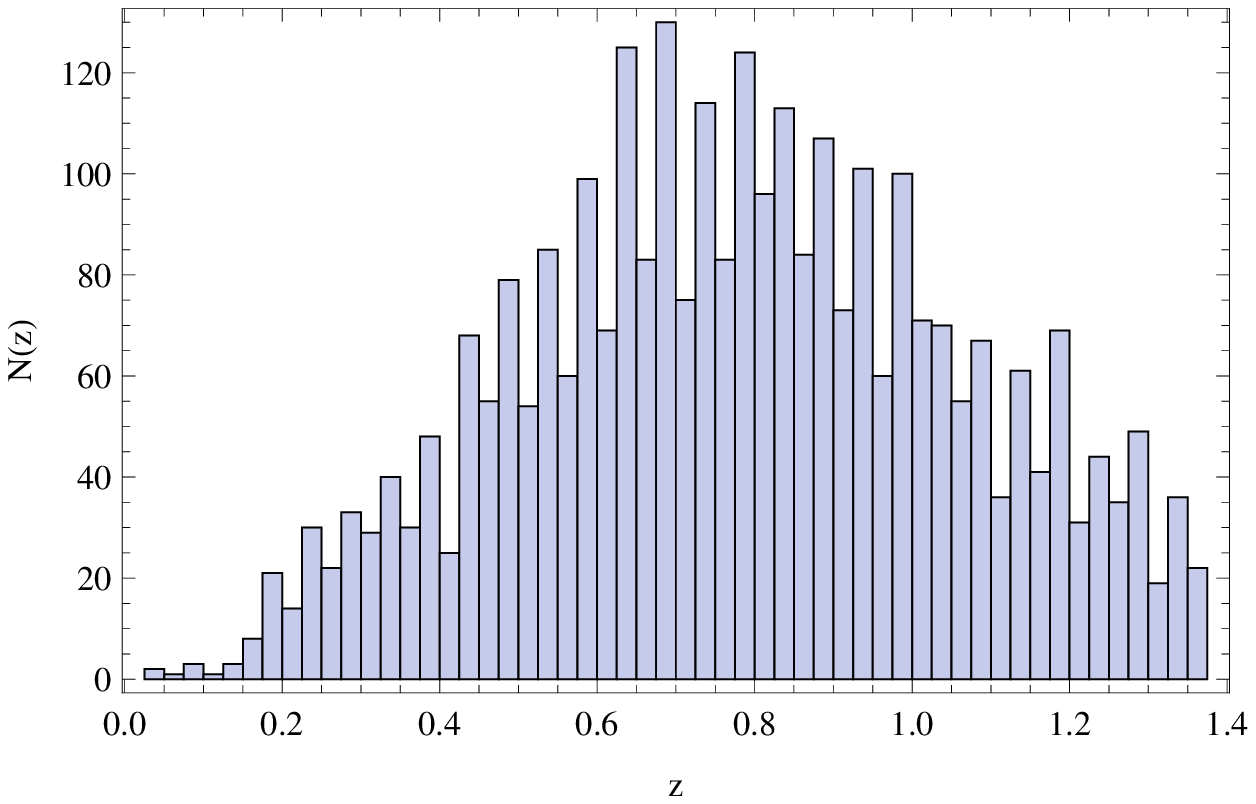}} \goodgap
\subfigure{\includegraphics[width=5cm]{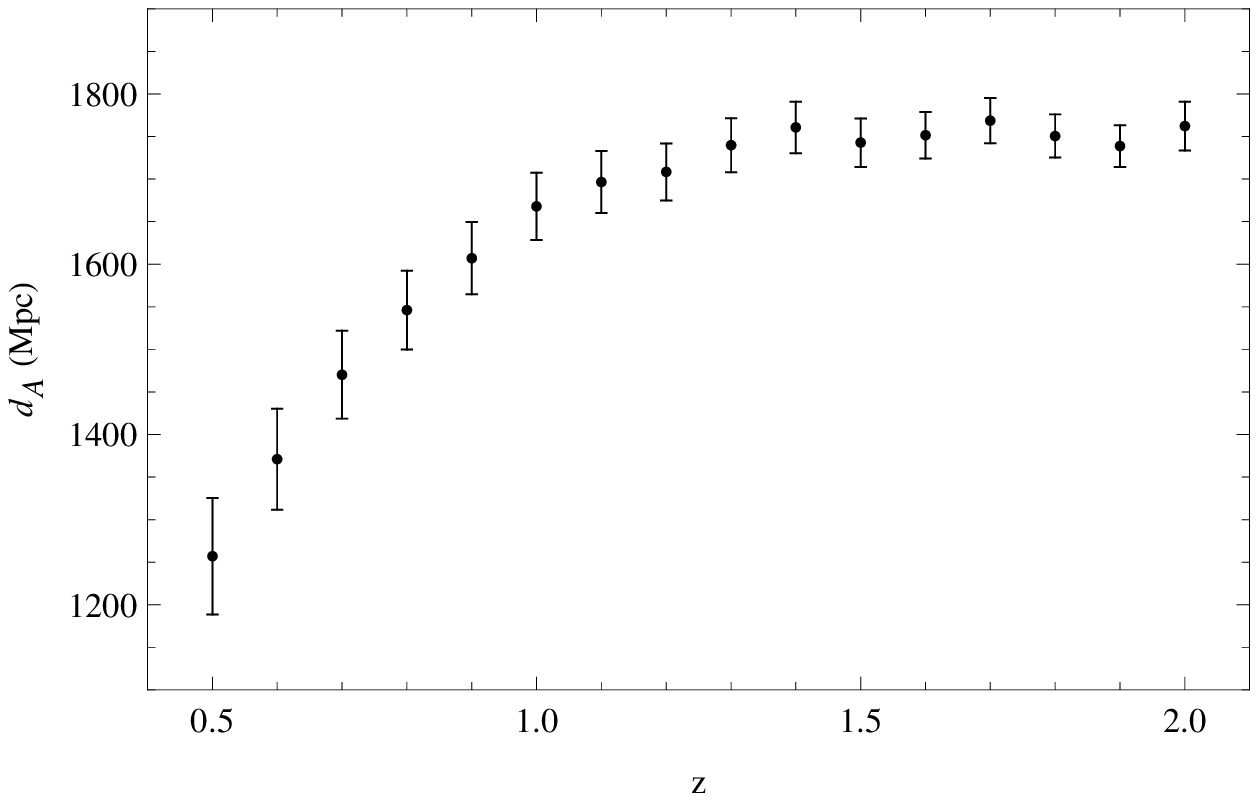}} \goodgap
\subfigure{\includegraphics[width=5cm]{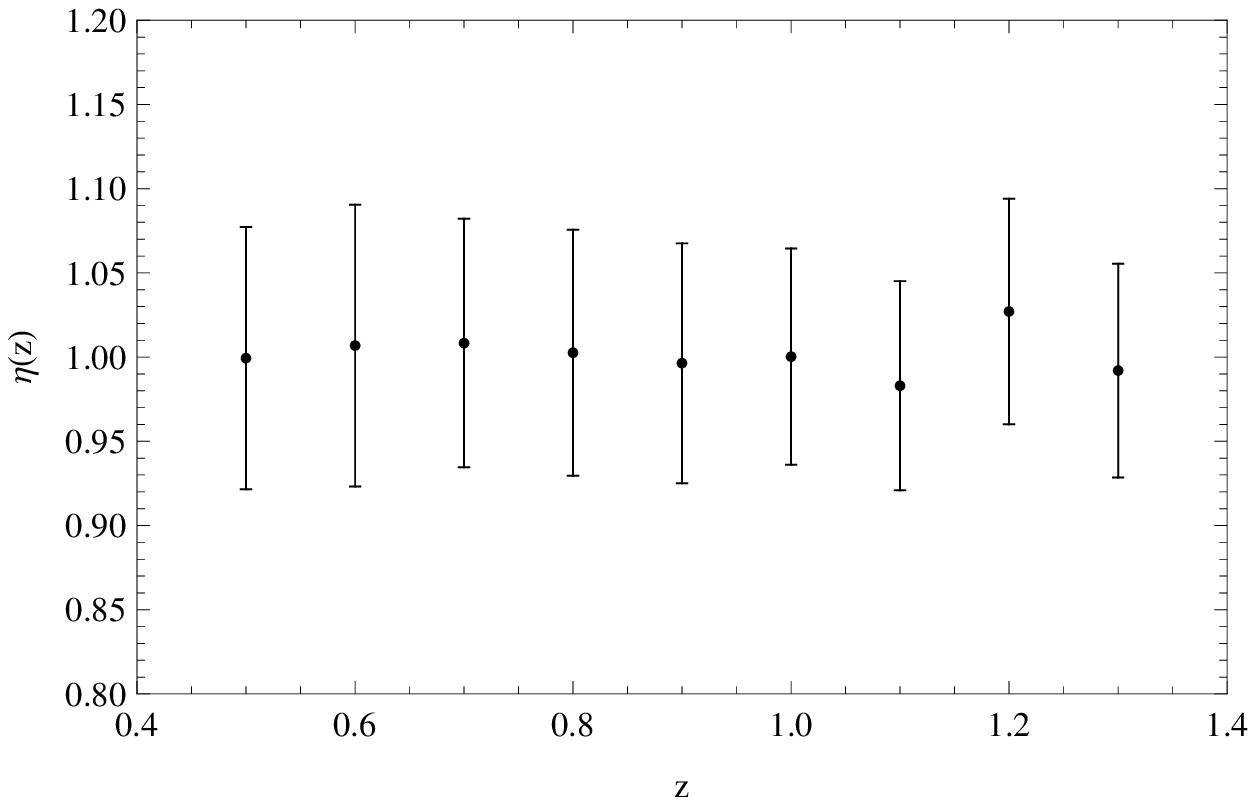}} \goodgap
\caption{Simulated data for an Euclid\,-\,like mission. {\it Left.} SNeIa redshift distribution (normalized so that the area under the histogram is 1). {\it Centre.} Angular diameter distance data. {\it Right.} Inferred $\eta(z)$ using local regression and the simulated SNeIa.}
\label{fig: simplot}
\end{figure*}

\section{DDR vs future data}

In order to escape the uncertainties on the cluster geometry, one must rely on a different tracer to estimate the ADD at a given redshift $z$. Baryon Acoustic Oscillations immediately stand out as ideal candidates to this aim. Indeed, the precise determination of the galaxy power spectrum as function of both the radial and tangential component of the wave vector allows to constrain $D_A(z_{med})/r_s$ and $H(z_{med}) r_s$, $r_s$ being the sound horizon, and $z_{med}$ the median redshift of the survey. Assuming that such a measurement is available, one can then rewrite the DDR in terms of the scaled ADD $\tilde{D}_A(z) = D_A(z)/r_s$ as\,:

\begin{displaymath}
\eta(z) = \frac{D_L(z) (1 + z)^{-2}}{D_A(z)} = \frac{D_L(z)}{\tilde{D}_A(z)} \frac{(1 + z)^{-2}}{r_s} \ .
\end{displaymath}
This can be conveniently rewritten as

\begin{equation}
r_s \eta(z) = \frac{D_L(z) (1 + z)^{-2}}{\tilde{D}_A(z)}
\label{eq: ddrbao}
\end{equation}
so that the rhs only contains observable quantities, while the lhs is a function of the sound horizon distance $r_s$ (which is a constant) and the parameters entering the adopted $\eta(z)$ phenomenological expression. Let us now suppose that a galaxy survey has the possibility to determine the power spectrum in ${\cal{N}}_{BAO}$ bins with sufficient accuracy to provide ${\cal{N}}_{BAO}$ measured values of the scaled ADD $\tilde{D}_A(z)$ with $z$ a characteristic redshift of the bin (e.g., the central or the median value). We can then resort to local regression on SNeIa to estimate the LD at the sampled $z$ and then get a catalog of $D_L(z) (1 + z)^2/\tilde{D}_A(z)$ measurements. This sample could then be fitted to an assumed $\eta(z)$ model, but this only gives constraints on $r_s \eta_0$ and $r_s \eta_a$. However, the sound horizon distance $r_s$ is well constrained by CMBR data in a (almost) model independent way and with an error which can be as small as $0.1\%$ according to what is forecasted for Planck. We can therefore assume that $r_s$ is known and directly use the ADD as $D_A(z) = r_s \tilde{D}_A(z)$ so that the same fitting analysis used with cluster data can be implemented for ADDs traced by the BAO.

The combination of BAOs to infer ADDs and local regression to estimate LD at the same ADD redshift allows us to get a set of measured values for $\eta(z)$ which is free from the two most problematic systematic errors that can mimic a deviation of the DDR even if such a violation of the Etherington reciprocity theorem should not be present at all. Unfortunately, while the available SNeIa samples are numerous enough to allow a decent reconstruction of $\mu(z)$ through the local regression method, present day BAOs measurements only allow to constrain $r_s/D_V(z)$, with $D_V(z) = [c z (1 + z)^2 D_A^2(z)/H(z)]^{1/3}$ the so called volume distance \cite{Eis05}. We have therefore to rely on future data to apply the test outlined above. Note that waiting for future data is a valid help also for improving LD estimates from SNeIa. Indeed, next to come SNeIa surveys will both increase the statistics and offer a better control of the systematics so that we can reduce the errors on the reconstructed LD thanks to both a larger subsample for each $z$ and outliers rejection.

\subsection{The simulated dataset}

In order to investigate the potential of combined BAO\,+\,SNeIa to constrain the DDR, we rely here on simulated data assuming an Euclid\,-\,like survey. Euclid \cite{Euclid} is a candidate ESA mission to map the geometry and the evolution of the dark universe to an unprecedented precision setting high accuracy constraints on dark matter, dark energy and modified gravity. To this end, two independent cosmological probes will be used, namely weak gravitational lensing and BAO, measuring the shape and the spectra of galaxies over $\sim 15000 \ {\rm deg^2}$ of extragalactic sky in both visible (down to $\sim 24.5$ AB mag in the visible wide R+I+Z filter) and NIR (up to 24 mag in Y, J, H filters), up to redshift $z \sim 2$. A deep survey (two magnitudes deeper than the wide) over a $40 \ {\rm deg^2}$ area will also be conducted for legacy science and could offer the possibility to detect $\sim 3000$ SNeIa. The possibility to both measure BAO and build up a SNeIa catalog makes Euclid an ideal tool to provide all the ingredients we need to check the validity of the DDR so that we use this mission as test case for our proposed method assuming the fiducial cosmological model described in Section II.

\subsubsection{SNeIa data}

Let us briefly describe how we simulate the SNeIa sample\footnote{The code we used has been developed to investigate how many SNe can be detected by an Euclid\,-\,like survey notwithstanding their type. As such, although we are only interested in SNeIa, we will automatically get for free also core collapse SNe.}. As a first step, we choose template light curves for each SN type (not only SNeIa, but also IIP, IIL, IIn and Ibc) as well as SN rates as function of redshift. Starting from the results of the LOSS \cite{Loss} survey for the magnitude peak and the Gaussian SN mag distribution, Montecarlo simulations are then performed generating artificial SNe (with expected total counts computed from the above template) and random redshifts and explosion epochs. Depending on the survey strategy, one can then compute the total number of SNe of each type which are detected at least one time and then impose some cuts on the number of epoch each SN is detected. Such cuts then allow to finally get the number of SNeIa which could be used for cosmology (i.e., that have a sufficiently well sampled lightcurve to determine their distance modulus) and their redshift distribution. To this end, we assume an observational strategy consisting in a first two months phase spent monitoring a $20 \ {\rm deg^2}$ field to a depth of $24.4 \ {\rm mag}$ at a 4 days cadence. This is immediately followed by a period with a 10 days cadence for 15 epochs to a depth of $24 \ {\rm mag}$. Then this same setup is repeated over a second $20 \ {\rm deg^2}$ patch of the sky thus finally giving us a sample of $3053$ SNeIa with $0.03 \le z \le 1.37$ and $z_{med} = 0.78$ with a redshift distribution plotted in the left panel of Fig.\,\ref{fig: simplot}. It is worth noting that the actual strategy that will be implemented by Euclid has still to be decided. We nevertheless stress that the expected SNeIa number is the same as what we are getting here so that we can confidently rely on our simulated dataset as a first guess of an Euclid\,-\,like catalog. To each SN in the sample, we estimate the error on the distance modulus as \cite{kim}\,:

\begin{equation}
\sigma_{\mu}(z) = \sqrt{\sigma_{sys}^2 + (z/z_{max})^2 \sigma_m^2}
\label{eq: sigmamusn}
\end{equation}
with $z_{max}$ the maximum redshift of the sample, $\sigma_{sys}$ an irreducible scatter and $\sigma_m$ depending on the photometric accuracy. Although these number have still to be computed for the Euclid SN survey\footnote{A different and more detailed strategy to forecast the precision on the distance modulus determination from the SN lightcurve has been described in \cite{Astier11}. We have preferred to not use their method since it involves a lot of further unknown parameters (such as the SALT2 color correction terms) thus introducing a degree of arbitrariness in the simulations that we prefer to avoid. It is worth noting, however, that they use a smaller value of $\sigma_{int}$ so that their uncertainties are likely smaller than our ones. As such, should their method turn out to be more reliable than our phenomenological formula, the results presented here would overestimate the impact of uncertainties thus leading to a conservative estimate of the final constraints on the DDR quantities.}, we set here $(z_{max}, \sigma_{sys}, \sigma_m) = (1.4, 0.15, 0.02)$ mimicking a typical space based survey. Denoting with $\mu_{fid}(z)$ the predicted value from our fiducial cosmological model, we then assign to each SN, a distance modulus randomly generated from a Gaussian distribution centred on $\mu_{fid}(z)$ and variance $\sigma_{\mu}(z)$ from Eq.(\ref{eq: sigmamusn}) above. The measurement error is finally set to $\sigma_{obs}(z) = [\sigma_{\mu}(z)/\mu_{fid}(z)] \mu_{obs}(z)$ thus finally obtaining the simulated SNeIa dataset we need as input to the local regression technique.

\subsubsection{BAO data}

We now discuss the simulated ADD measurements from BAO. To this end, we use the method developed and tested in \cite{SE07} to forecast the percentage error on $D_A(z)/r_s$ from a BAO survey as a function of both the fiducial cosmological model and the survey characteristics. To this end, it is worth first remembering that Euclid will perform slitless spectroscopy  for galaxies with an H$\alpha$ flux down to $f \simeq 4 \times 10^{-16} \ {\rm erg \ s^{-1} \ cm^{-2}}$ so that its main target will be star forming galaxies. Such an information is important to both estimate the redshift number distribution of detectable sources and the linear bias to be applied to match the matter and galaxy power spectra. Following \cite{melita}, we will assume a $20000 \ {\rm deg^2}$ survey over the redshift range $0.5 \le z \le 2$ with $dN/dz$ obtained by multiplying the one in \cite{Geach} by a success rate $\epsilon = 0.35$ for a conservative choice, while the linear bias varies with the redshift according to the model in \cite{Orsi}. Different from \cite{melita}, we consider 16 equally spaced redshift bins with bin width $\Delta z = 0.1$ in order to increase the number of $D_A(z)$ measurements, but we stress that we can actually estimate $\eta(z)$ only for the first 9 bins since the SNeIa sample does not extend to $z > 1.3$ so that no LD determination is available for the higher redshift bins.

Two further ingredients are needed before using the \cite{SE07} code. First, one has to set the spectral index of scalar perturbations, denoted as $n_s$, and the variance of density perturbations in a sphere of radius $8 h^{-1} \ {\rm Mpc}$, usually referred to as $\sigma_8$. In agreement with the WMAP/ results, we choose $(n_s, \sigma_8) = (0.96, 0.809)$. Finally, in order to avoid the problem of modeling nonlinear effects, we cut the power spectrum to a maximum wavenumber $k_{max}$ determined by solving $\sigma^2(1/k_{max}, z) = 0.25$, with $\sigma^2(R, z)$ the variance over the scale $R = 1/k$ for the power spectrum at redshift $z$. Note that this leads to a redshift dependent upper limit on the usable power spectrum, although a conservative good approximation is to set $k_{max} \sim 0.15 h \ {\rm Mpc^{-1}}$ independent on $z$.

The code then outputs $\sigma_{s_{\perp}}$, i.e., the error on $\ln{D_A(z)/r_s}$ so that, if we assume that the error on $r_s$ is negligible, we simply get $\sigma_{D_A}/D_A = \sigma_{s_{\perp}}$. As a simplifying (but yet realistic assumption), we will associate this error to the ADD measurement at the centre of the redshift bin. We then generate $D_A(z)$ from a Gaussian distribution centred on the fiducial ADD and with variance equal to the one outputted from the code and finally scale the measurement error according to the ratio between the simulated and fiducial distance. The data thus generated are shown in the central panel of Fig.\,\ref{fig: simplot}, while the right panel plots the inferred $\eta(z)$ measurements using the local regression technique to estimate the LD for the BAO ADDs measurements (up to $z = 1.3$).

\subsection{Constraints on DDR parameters}

The above simulated dataset are input to the same fitting procedure analysis we have used for the present day data. We start by discussing the results for one representative realization of the SNeIa and BAO data. For the one parameter models (i.e., with $\eta_0 = 1$), we get\footnote{Since we are dealing with simulated datasets, the best fit values have no particular meaning so that we could also report only the $1 \sigma$ uncertainties. We have nevertheless preferred to give also the best fit $(\eta_0, \eta_a)$ in order to show that there is no bias induced by the simulations and the fitting method.}\,:

\begin{displaymath}
\eta_a = 0.001 \pm 0.067
\end{displaymath}
for the first case in Eq.(\ref{eq: etapar}), and

\begin{displaymath}
\eta_a = 0.001 \pm 0.047
\end{displaymath}
for the second ansatz. A comparison with the results for the simulated case using local regression discussed at the end of Section IIIA shows that, although we now use a smaller dataset (only 9 instead of 38 points), the errors on $\eta_a$ have been reduced by a factor two. Such a large reduction is a consequence of two effects. On one hand, the increased size of the SNeIa sample (by a factor ten) allows to have more points in each of the local bins used to fit the low order polynomial used in the local regression method thus reducing the error on $D_L(z)$. On the other hand, BAO data allows to measure $D_A(z)$ with an accuracy of order $5\%$ so that the final uncertainty on $\eta(z)$ is quite small. As a result, the lower statistics offered by this method is more than compensated by the far better precision thus shrinking the $\eta_a$ confidence ranges.

When $\eta_0$ is left free, we find\,:

\begin{displaymath}
\eta_0 = 0.994 \pm 0.180 \ \ , \ \ \eta_a = -0.016 \pm 0.321 \ \ ,
\end{displaymath}

\begin{displaymath}
\eta_0 = 0.940 \pm 0.175 \ \ , \ \ \eta_a = -0.009 \pm 0.173 \ \ ,
\end{displaymath}
for the two models in Eq.(\ref{eq: etapar}). Compared to the present day simulated data, we now find that the constraints on $\eta_0$ are actually poorer, while the opposite result is obtained, instead, for the $\eta_a$ parameters whose confidence ranges are smaller. While the first result is a consequence of the lower statistics which is no more compensated by the increased precision because of the presence of two parameters to fit, the improvement in the $\eta_a$ constraints is related to the larger redshift range probed by the BAO data. It is, however, worth stressing that the statistical uncertainties on $(\eta_0, \eta_a)$ coming out from the fit are actually not the only source of error. As we have seen when fitting the present day data, systematic errors can also be larger than the statistical ones and bias the inferred best fit values. From this point of view, the BAO method is free from this problem so that should be preferred over the clusters as an ADD tracer.

Finally, we check whether the method used is able to recover the input parameters. To this end, we have run $\sim 100$ realizations of the SNeIa and BAO future data and repeated the fit for each of them. For the one parameter models, averaging the median $\eta_a$ over the full set of simulations, we find\,:

\begin{displaymath}
\langle \eta_a \rangle = -0.001 \pm 0.004 \ \ , \ \ \langle \eta_a \rangle = -0.001 \pm 0.003 \ \ ,
\end{displaymath}
for the linear and logarithmic $\eta(z)$ ansatz, respectively, and where the error is the standard deviation of the (approximately) Gaussian distribution of the results. Leaving $\eta_0$ as a free parameter, we get\,:

\begin{displaymath}
\langle \eta_0 \rangle = 0.98 \pm 0.02 \ \ , \ \ \langle \eta_a \rangle = -0.01 \pm 0.03 \ \ ,
\end{displaymath}
for the linear model, and 

\begin{displaymath}
\langle \eta_0 \rangle = 0.94 \pm 0.01 \ \ , \ \ \langle \eta_a \rangle = 0.00 \pm 0.02 \ \ ,
\end{displaymath}
for the logarithmic one. Such results suggest that the median $\eta_a$ value outputted from the fit is on average consistent with the input one for both the linear and logarithmic model independent on the use of the $\eta_0 = 1$ assumption. On the contrary, $\eta_0$ is less well recovered because of the degeneracy with the nuisance $\Delta_0$ parameter. Although this could add a note of caution in using the proposed method, it is nevertheless worth stressing that, for all the simulations, the statistical error on $\eta_0$ is roughly the same as the one reported above for the representative case. As a consequence, the value $\eta_0 = 1$ is always well within the $1 \sigma$ error so that we conclude that the bias is not statistically meaningful.

\section{Conclusions}

It is common to say that we are living in the era of precision cosmology. While this is only partly true today, one can be confident that future data will us make enter an epoch where we can not only improve the precision on the constraints on a given cosmological model, but also test the cornerstones of observational cosmology. Although its importance is usually underrated, the Etherington reciprocity law stands out as one of the fundamental pillars our interpretation of astrophysical data is based on. Next to come surveys will have the sufficient quality to promote the distance duality relation (which is the most used incarnation of the Etherington law) from an {\it a priori theoretical assumption} to the rank of a {\it relation which can be observationally validated}.

In order to test the validity of the DDR, one needs to trace both the luminosity and angular diameter distance for a set of redshift values. We have here followed the usual approach relying on clusters data to estimate the ADD and SNeIa as LD tracer. We have, however, improved the standard analysis by introducing the local regression technique to avoid the redshift mismatch problem (i.e., the difference between the redshift of the cluster and those of the SNeIa used to infer the corresponding LD). This simple and widely tested method allows to strongly reduce the bias on the $\eta(z)$ parameters thus increasing the reliability of the constraints and hence the test of the DDR validity. Unfortunately, the poor quality of the cluster ADDs determination still leads to large confidence ranges preventing to draw any statistically meaningful conclusion on the violation of the DDR over the redshift range probed by the available data. Moreover, the results strongly depend on the assumptions on the cluster geometry so that one should first find a method to correct for this effect or propagate this uncertainty on the final error on the $(\eta_0, \eta_a)$ parameters introduced to quantitatively check the DDR validity.

In an attempt to escape this problem, we have here proposed to use BAO as an alternative ADD tracer. Being the physics of BAO well understood, the systematics connected with this method can be easily quantified and satisfactorily corrected for with future galaxy surveys data. Since the present day data are too poor to implement this test in an efficient way, we have relied on simulated samples of both BAO ADD measurements and SNeIa distance moduli determinations considering a fiducial Euclid mission as source of both datasets. Such an analysis has highlighted the virtues of the proposed approach showing that the error bars are halved if one forces $\eta_0 = 1$. When this assumption is abandoned, we find only a modest decrease of the relative uncertainty on $\eta_0$ with respect to present day data, but the constraints on $\eta_a$ are still strengthened by a factor two. Moreover, the lack of systematic errors makes this approach highly preferable over the use of cluster data as ADD tracers.

It is worth noting that the proposed approach does not exploit the full potential of BAO. Indeed, while BAO allows to determine the ADD up to redshift $z = 2$, the quantity $\eta(z)$ can only be determined up to $z = 1.4$, this latter being the maximum redshift available tested by the SNeIa Hubble diagram. In order to push further this limit, one could rely on a different SNeIa survey able to detect a statistically meaningful number of objects at $z > 1.4$ with sufficient precision. As an alternative approach, one should find a different LD tracer. Gamma ray bursts (GRBs) stand out as ideal candidates from this point of view since they can be detected up to $z \sim 8$ \cite{Salvaterra2009} thanks to the huge energy released during the explosion. Unfortunately, the use of GRBs as standardizable candles is still in its infancy so that, notwithstanding the first released GRBs Hubble diagrams \cite{CCD09,S07}, a careful analysis of the systematics has still to be fully done (but see, e.g., \cite{marcy} for recent encouraging results). Should future data validate the GRBs as LD tracer, one could use them as input to the local regression technique and trace $\eta(z)$ over the full redshift range probed by BAO surveys.

As a final remark, it is worth noting that the proposed method will allow not only to check the foundations of observational cosmology by giving an empirical validation of the universally assumed Etherington law, but also open the way to completely new physics should this test find out a statistically meaningful violation of the DDR.

\acknowledgements

VFC and SS are funded by the Italian Space Agency (ASI) through contract Euclid\,-\,IC (I/031/10/0). We thank E. Cappellaro for help with developing the SNeIa simulation code and T. Kitching and M. Kunz for comments on an earlier version of the manuscript.

\end{document}